\documentclass[traditabstract]{aa}
\usepackage{txfonts}
\usepackage{natbib}
\bibpunct{(}{)}{;}{a}{}{,}

\usepackage{epsfig}
\usepackage{graphicx}
    \DeclareGraphicsExtensions{.eps, .jpg}
\usepackage{color}

\newcommand{\Msun}{M_{\odot}}
\newcommand{\Mdot}{\dot{M}}
\newcommand{\MdotEdd}{\dot{M}_{\rm Edd}}

\newcommand{\be}{\begin{equation}}                                 
\newcommand{\ee}{\end{equation}}                                   
\newcommand{\bea}{\begin{eqnarray}}                                
\newcommand{\eea}{\end{eqnarray}}                                  
\newcommand{\nn}{\nonumber}                                                   
\definecolor{gray}{rgb}{.6,.6,.6}                                  %
\definecolor{green}{rgb}{0,.6,0}                                   %
\definecolor{red}{rgb}{0.6,0,0}                                    %

\begin{document}

%
\title{Modelling the black hole silhouette in Sgr~A* with ion tori}
%
\author
{
Odele~Straub \inst{1}
\and
Frederic~H.~Vincent \inst{2,3}
\and
Marek~A.~Abramowicz \inst{1,4,5}
\and
Eric~Gourgoulhon \inst{3}
\and
Thibaut~Paumard \inst{2}
}
\institute{
Nicolaus Copernicus Astronomical Center, Bartycka 18,
00-716 Warsaw, Poland
\and
LESIA, CNRS UMR 8109, Observatoire de Paris, Universit\'e Pierre et Marie Curie, Universit\'e Paris Diderot, 
92190 Meudon, France
\and
LUTH, CNRS UMR 8102, Observatoire de Paris, Universit\'e Paris Diderot, 
92190 Meudon, France
\and
Department of Physics, G\"oteborg University,
SE-412-96 G\"oteborg, Sweden
\and
Institute of Physics, Silesian University, 746-01 Opava, Czech Republic
}
\date{Received XXXX; accepted XXXX }
\abstract{
We calculate the ``observed at infinity'' image and spectrum of the accretion structure in Sgr~A*, by modelling it as an optically thin, constant angular momentum ion torus in hydrodynamic equilibrium. The physics we consider includes a two-temperature plasma, a toroidal magnetic field, as well as radiative cooling by bremsstrahlung, synchrotron and inverse Compton processes. Our relativistic model has the virtue of being fully analytic and very simple, depending only on eight tunable parameters: the black hole spin and the inclination of the spin axis to our line of sight, the torus angular momentum, the polytropic index, the magnetic to total pressure ratio, the central values of density and electron temperature and the ratio of electron to ion temperatures. The observed image and spectrum are calculated numerically using the ray-tracing code GYOTO. Our results demonstrate that the ion torus model is able to account for the main features of the accretion structure surrounding Sgr~A*.
}
\authorrunning{O.\,Straub et al.}
\titlerunning{Imaging ion tori around Sgr~A*}
  \maketitle

\section{Introduction}
\label{sec:intro}
The Galactic Centre is one of the most interesting regions for scientific investigation because it is the closest available galactic nucleus and, therefore, can be studied with resolutions that are impossible to achieve with other galaxies. The radio source Sagittarius A* (henceforth Sgr~A*) is associated with the super-massive black hole at the centre of our Galaxy. Its mass, $M=4.31 \times 10^6 \Msun$, has been determined in particular from the complete and highly eccentric orbit of the star $\boldmath{S2}$, which passes Sgr~A* in its pericentre at a mere 17 light-hours, i.e. $1400\; r_{\rm S}$, where $r_{\rm S} \equiv 2 GM/c^{2}$ is the Schwarzschild (or gravitational) radius \citep{sch+02, bow+04, gil+09}. Given the Sgr~A* distance of 8.33 kpc, the Schwarzschild radius corresponds to an angular size of around $10\,\mu$as in the sky, making the Galactic Centre black hole an ideal candidate for near-future microarcsecond interferometric technologies \citep{pau+08,eis+08,eis+10,bro+11b}.

In this context, several authors have calculated images of various theoretical models of accretion structures around Sgr~A*. Numerical simulations reach from three-dimensional non-relativistic MHD \citep{gol+05, hua+07} to two-dimensional GRMHD \citep{mos+09, hil+10} and three-dimensional GRMHD \citep{dex+10, dib+12}. The size and shape of the ``black hole silhouette'' cast by the black hole on the accretion structure, is determined by the photon orbit and depends thus only on black hole mass and spin. Therefore, with the known black hole mass in the Galactic Centre the spin could be estimated from fitting the calculated size and shape of the shadow to the size and shape observed. The looks (spatial extent, shape, brightness distribution, etc.) of the entire image, however, depend not only on black hole mass and spin but also on the details of the accretion structure around Sgr~A*. Most of them are uncertain, e.g., the chemical abundance, the radiative processes, the inclination. They have not been sufficiently examined yet, and a wide range of the relevant parameter space remains unexplored.

The accretion structure around Sgr~A* is most probably a radiatively inefficient advection dominated accretion flow (ADAF), as \citet{nym95} rather convincingly demonstrated by means of spectral fitting. In the following, an approximate analytic torus model associated to the family of Polish doughnuts introduced by \citet{abr+78} and \citet{jar+80} is constructed for a radiatively inefficient accretion flow to describe the accretion structure in Sgr~A*. The model depends on a few tunable parameters which can be changed and adjusted so that they can cover the whole parameter space of the problem. The general model of the source, and the ray-tracing tailored especially for it, may as well be applied to spectral calculations of other black hole candidates, both optically thin (ion tori, to model e.g. some spectral states in X-ray binaries) and optically thick (Polish doughnuts, to model e.g. some luminous AGNs and ULXs).

\citet{sha+76} designed a hot, two-temperature disc model to describe the strong X-ray emission observed in Cygnus X-1, which was too hot to be understood in terms of the standard model. They found an optically and geometrically thin solution branch in which ions and electrons are weakly coupled, having different temperatures and being in energy balance. This solution, however, is thermally unstable. 

Ion tori were proposed independently by \citet{ich77} (though not by this name) and \citet{ree+82} to model a gas flow that emits little detectable radiation but is at the same time able to power radio jets, both in galactic sources and microquasars. Ion tori are geometrically thick, gas (ion) pressure supported spheroidal structures, located at the inner regions of accretion flows. They are extremely optically thin and very radiatively inefficient. These ion tori consist of a fully ionised plasma, hence their name, in which protons and electrons are thermally decoupled (each following its own temperature distribution) and not in local energy balance. They are assumed to be low-$\alpha$ accretion flows threaded with magnetic fields, so that the ordinary molecular viscosity is suppressed in favour of magnetic/turbulent viscosity\footnote{The $\alpha$-prescription is applied also in this context.}. Due to their vertically extended shape, $H/r \lesssim 1$, ion tori naturally create a pair of funnels through which magnetic flux can collimate and escape. Ion tori only exist in the sub-Eddington accretion regime, $\Mdot < \Mdot_{\rm crit} < \MdotEdd$ \citep[see e.g.][]{ich77, ree+82}, where radiative cooling is inefficient enough that ions stay sufficiently hot (near virial) during the whole inflow time, i.e., $t_{\rm diff} \gg t_{\rm accr}$. Otherwise, at higher mass accretion rates, cooling causes the puffed up inner spheroidal region to gradually become opaque and deflate to a standard thin disc. As a consequence of $t_{\rm diff} \gg t_{\rm accr}$, the ions carry most of the energy with them into the black hole. Instead of ``the ions are not cooling down fast enough'' one may also and equivalently say that the disc is locally ``advectively cooled'' because energy is carried away by ions. Therefore, this branch of solutions was renamed to ``Advection Dominated Accretion Flow'' (ADAF) \citep[see e.g.][]{ny94, ny95, abr+95, esi+97}.
 
In this work, we model Sgr~A* by ion tori very much similar to these described by \citet{ree+82}. They are based on the mathematical description introduced by Paczy{\'n}ski and collaborators in their studies of {\it Polish doughnuts} \citep{abr+78, jar+80}. Although Polish doughnuts have some of their important characteristics directly {\it opposite} to these of ion tori (they are optically thick, radiation pressure supported and correspond to super-Eddington accretion rates, ${\dot M} \gg {\dot M}_{\rm Edd}$) both classes of tori have the same equipotential structure and very similar dynamical properties. In particular, they both have the ``Roche lobe'', i.e. a critical equipotential that crosses itself along the ``cusp'' at $r = r_{\rm cusp}$ (see Fig.~\ref{fig:equipotential}). Roche lobe overflow causes the {\it dynamical} mass loss from the torus to the black hole, with no need of help from viscosity. Thus, the accretion flow at radii $r \le r_{\rm cusp}$ is regulated by the black hole strong gravity and not by viscus processes. The cusp should be regarded as the {\it inner edge of the disk}; for small accretion rates it corresponds to ISCO, for higher accretion rates the cusp locates closer to the black hole. Both ion tori and Polish doughnuts are dynamically unstable with respect to the \citet{pap+84} instability. However, as proved by \citet{bla87}, this instability is suppressed by the Roche lobe overflow \citep[see also][]{nar+93}.

These properties, shared by ion tori and Polish doughnuts, are genuine and robust also for much more general toroidal structures around black holes. In particular, they do not depend on the angular momentum distribution inside the torus. It is customary to assume that angular momentum is constant, $\ell = \ell_0 =$ const, as this leads to {\it remarkably simple} final analytic formulae. This is an assumption adopted here. In follow-up papers we will relax this assumption and calculate observed properties of ion tori with $\ell \not=$ const. Dynamical models for such tori have been calculated e.g. by \citet{qia+09}.

The paper is structured as follows. In section \ref{sec:tori} we construct the hydrodynamical torus and in section \ref{sec:physics} and section \ref{sec:ray-tracing} we discuss the radiative properties. Section \ref{sec:results} shows the resulting spectra in relation to broadband data of Sgr~A* and a series of images. The presented spectra are neither chi-squared fits nor best guesses, they merely illustrate the performance of the ion torus model. Section \ref{sec:discussion} presents the conclusions.

\section{The geometry of fat tori}
\label{sec:tori}

\subsection{Kerr metric} 
In Boyer-Lindquist spherical coordinates $(t, r, \theta, \phi)$, geometrical units $c = 1 = G$, and signature $(-, +, +, +)$, the Kerr metric 
line element has the form:
\bea
  \mathrm{d}s^2 &=&
  -\left(1-\frac{2 M r}{\Xi}\right) \,\mathrm{d}t^2 
  - \frac{4 M r a }{\Xi} \sin^2\theta\,\mathrm{d}t\, \mathrm{d}\phi 
  + \frac{\Xi}{\Delta}\, \mathrm{d}r^2  \nn \\
&& 
  + \Xi \,\mathrm{d}\theta^2
  + \left(r^2 + a^2 +\frac{2 M r a^2 \sin^2\theta}{\Xi}\right)\sin^2\theta\, \mathrm{d}\phi^2,
\label{eq:kerr2}
\index{Kerr metric}
\eea
where $M$ is the black hole mass, $a\equiv J/M$ its reduced angular momentum ($J$ being the black hole angular momentum), $\Xi \equiv r^2 + a^2 \cos^2\theta$ and $\Delta \equiv r^2 - 2 M r + a^2$. 

Circular orbits in Kerr spacetime are obtained from the timelike geodesics of the metric (\ref{eq:kerr2}) for which the 4-velocity takes the form
$u^\mu = (u^t,0,0,u^\phi)$. 
The two constants of geodesic motion, associated respectively with stationarity and axisymmetry, are then expressible as 
\bea
  \mathcal{E} &=& - u_t = - u^t \,(g_{tt}+\Omega \,g_{t\phi}), \label{eq:E_u} \\
  \mathcal{L} &=& u_{\phi} = u^t\, (g_{t\phi}+\Omega \,g_{\phi\phi}) ,\label{eq:L_u}
\eea
where 
\be
  \Omega \equiv \frac{u^\phi}{u^t} 
\ee
is the angular velocity with respect to a distant observer. The specific angular momentum is defined by 
\be
  \ell \equiv \frac{\mathcal{L}}{\mathcal{E}} = - \frac{u_\phi}{u_t}.
\label{eq:def_ell}
\ee
At circular {\it geodesic} orbits, the angular momentum has its ``Keplerian'' form \citep{bar+72}:
\be
  \ell_{\rm K}(r,a) = \frac{M^{1/2}\,\left(r^2-2aM^{1/2}r^{1/2}+a^2\right)}{r^{3/2}-2Mr^{1/2}+aM^{1/2}}.
\label{eq:ell_K}
\ee
Here and in the remainder of the article, we consider $M$ a fixed parameter, so that we can take $\ell_{\rm K}$ to be a function of $(r,a)$ only. 

Important circular orbits in the Kerr metric are the marginally stable and marginally bound ones, corresponding to the following values of $r$ \citep{bar+72}: 
\bea 
  r_{\rm ms}(a) &=& M\,\left[3 + z_2 -((3 - z_1) (3 + z_1 + 2z_2))^{1/2}\right],\\
  r_{\rm mb}(a) &=& 2M - a +2M^{1/2}(M-a)^{1/2}. 
\label{marginally}
\eea
where $z_1 \equiv 1 + (1 - a^2/M^{2})^{1/3} \left[(1 + a/M)^{1/3} + (1 - a/M)^{1/3}\right]$ and $z_2 \equiv (3a^2/M^{2} + z_1^2)^{1/2}$. On these special orbits, the specific angular momentum takes the values
\be 
  \ell_{\rm ms}(a) = \ell_{\rm K}(r_{\rm ms}(a), a) ~~~\textnormal{and}~~~ 
  \ell_{\rm mb}(a) = \ell_{\rm K}(r_{\rm mb}(a), a).
\label{eq:l_ms_l_mb}
\ee

\begin{figure*}
\includegraphics[width=0.48\textwidth]{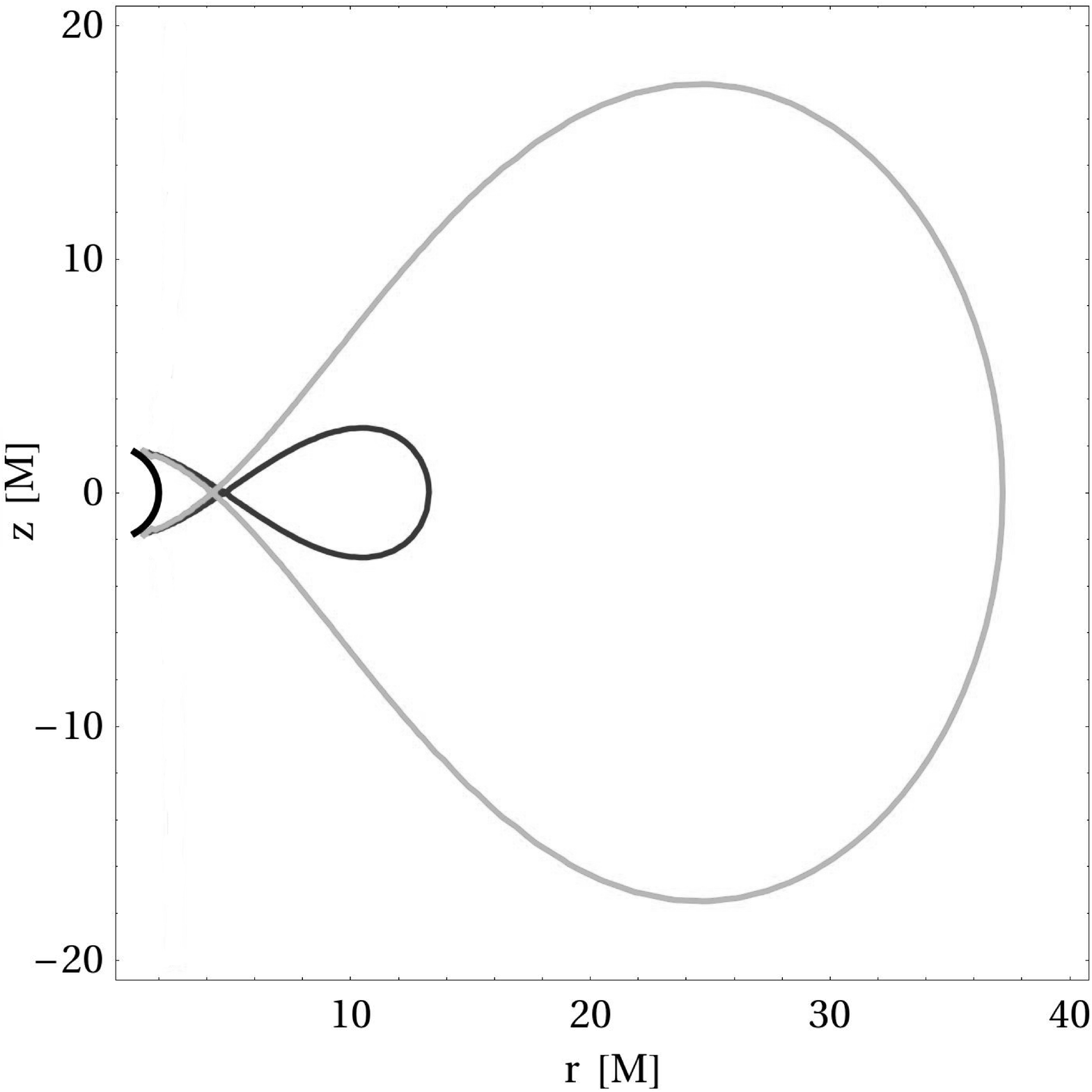}
\hfill
\includegraphics[width=0.48\textwidth]{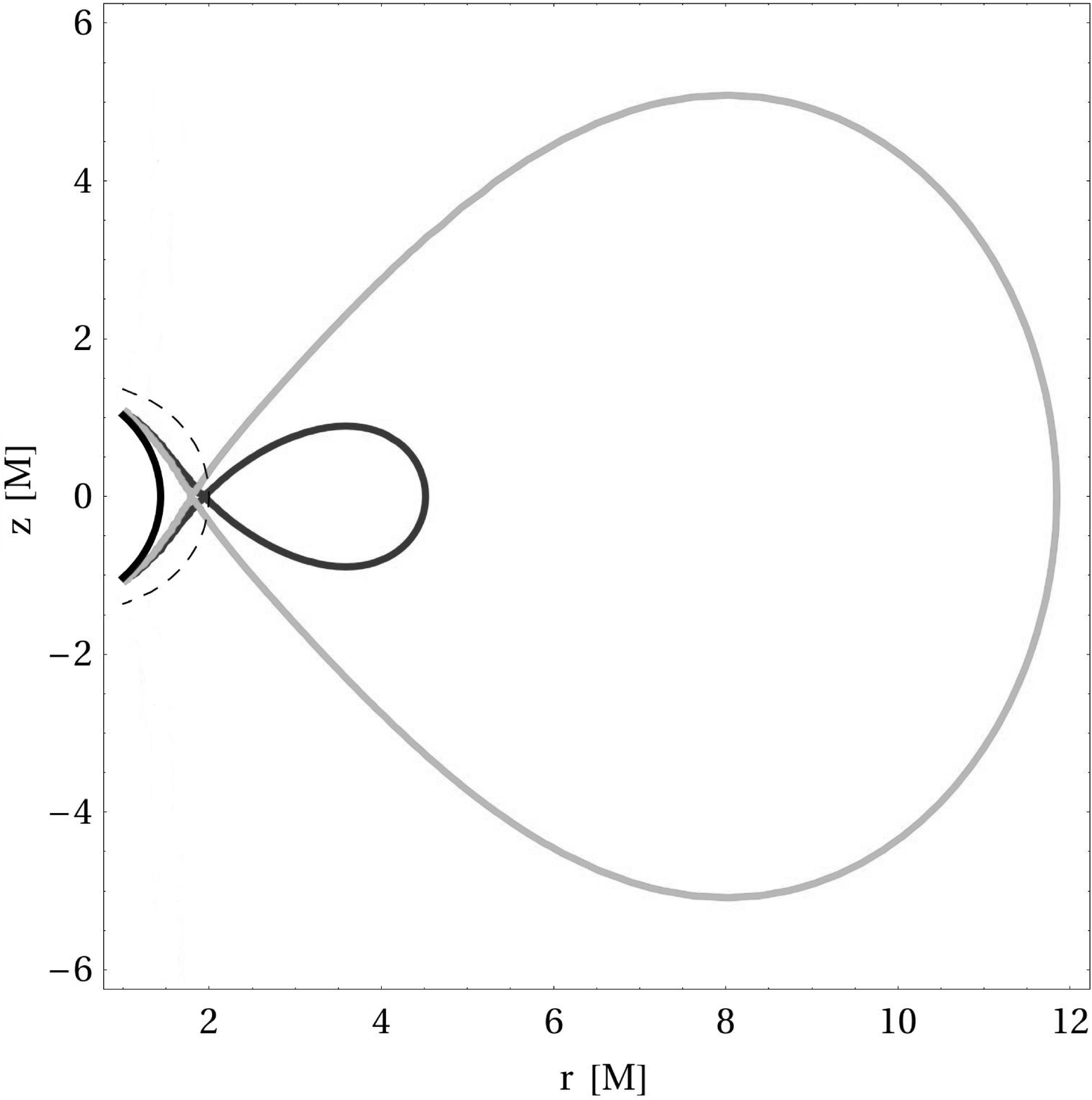}
\caption{Meridional cut through a $\lambda = 0.3$ (black curve) and a $\lambda = 0.7$ torus (grey curve) around a black hole (the black semi-circle is the event horizon). {\it Left panel:} Schwarzschild case. {\it Right panel:} Kerr case with $a = 0.9\,M$, the long-dashed semi-circle marking the ergosphere.}
\label{fig:equipotential}
\end{figure*}
%

\subsection{Fluid torus of constant specific angular momentum} 
\label{s:fluid_torus}

We consider a fluid torus of negligible self-gravitation around a Kerr black hole. Modelling the fluid as a perfect one, the stress-energy tensor is
\be
  T_{\mu\nu} = (\epsilon + P) u_\mu u_\nu + P g_{\mu\nu} , 
\ee
where $P$ is the fluid pressure and $\epsilon$ the fluid proper energy density. For a purely circular motion ($u^\mu = u^t(1,0,0,\Omega)$), it can be shown that the energy-momentum conservation equation, $\nabla_\nu T^\nu_{\ \, \mu} = 0$, takes the form 
\be
 \frac{\nabla_{\mu} P}{\epsilon+P} = -\nabla_{\mu} \ln(-u_t) + \frac{\Omega\nabla_{\mu}\ell}{1 - \Omega \ell},
\label{eq:euler full}
\ee
with $\ell$ related to the fluid 4-velocity components by Eq.~(\ref{eq:def_ell}). Assuming that the fluid is barotropic, $P=P(\epsilon)$, we introduce the enthalpy function 
\be 
  H \equiv \int_0^P \frac{\mathrm{d}P}{\epsilon+P} ,
\label{eq:def_enthalpy}
\ee
so that the left-hand side of Eq.~(\ref{eq:euler full}) becomes simply $\nabla_\mu H$. Following \citet{abr+78}, we consider models for which $\ell$ is constant within all the torus:
\be
  \ell= \ell_0.
\ee
Accordingly, the equation of motion~(\ref{eq:euler full}) reduces to $\nabla_\mu H = - \nabla_\mu \ln(-u_t)$ yielding
\be 
  H = W + \mathrm{const} , 
\label{eq:first_integ}
\ee
where we have introduced the potential
\be
  W \equiv - \ln(-u_t) . 
\ee
From the normalisation relation $u_\mu u^\mu = -1$, we get
\be
 W(r,\theta) = \frac{1}{2}\ln\left[ - \frac{g_{tt} + 2\Omega g_{t\phi} + \Omega^2 g_{\phi\phi}}{(g_{tt} + \Omega g_{t\phi})^{2}} \right].
\label{eq:W} 
\ee
where $\Omega$ should be considered as the following function of $(r,\theta)$:
\be 
  \Omega(r,\theta) = -\frac{g_{t\phi} + \ell_0 g_{tt}}{g_{\phi\phi} + \ell_0 g_{t\phi}} . 
\label{eq:Omega_ell0}
\ee
This last relation is easily derived by combining Eqs.~(\ref{eq:E_u}), (\ref{eq:L_u}) and (\ref{eq:def_ell}).

\citet{abr+78} proved that the cusp location should be between the mariginaly stable and the mariginaly bound orbit, which implies that $\ell_0$ must obey,
\be
 \ell_{\rm ms}(a) < \ell_0 < \ell_{\rm mb}(a), 
\label{eq:condl0}
\ee
with $\ell_{\rm ms}(a)$ and $\ell_{\rm mb}(a)$ given by Eq.~(\ref{eq:l_ms_l_mb}). Recast in terms of the dimensionless parameter
\be 
\lambda \equiv \frac{\ell_0 - \ell_{\rm ms}(a)}{\ell_{\rm mb}(a) - \ell_{\rm ms}(a)},
\label{eq:def_lambda}
\ee
the condition (\ref{eq:condl0}) is equivalent to 
\be
0 \le \lambda \le 1.
\ee
Given the radial dependence (\ref{eq:ell_K}) of the Keplerian specific angular momentum, which takes its minimum at $r=r_{\rm ms}(a)$, and the constraint (\ref{eq:condl0}), there are two values of $r$, $r_{\rm in}$ and $r_{\rm c}$ say, for which $\ell_{\rm K}(r,a) = \ell_0$. One has necessarily
\be
r_{\rm mb}(a) < r_{\rm in} < r_{\rm ms}(a) < r_{\rm c} . 
\ee
Since for $r=r_{\rm in}$ or $r = r_{\rm c}$, the actual angular momentum $\ell_0$ is equal to the Keplerian one, the gravity and centrifugal forces balance each other, implying $\nabla_\mu W = 0$. From the first integral (\ref{eq:first_integ}), we have then $\nabla_\mu H = 0$, or equivalently, via (\ref{eq:def_enthalpy}), $\nabla_\mu P=0$. A point where the gradient vanishes can be either a singular point (self-crossing of an equipotential) or an extremum. In the present case, $r=r_{\rm in}$ is the location where one of the equipotentials of $W$, the so-called \emph{Roche lobe}, self-crosses, as shown in Fig.~\ref{fig:equipotential}. This gives rise to a cusp at the torus surface \citep{abr+78}. On the other side, $r=r_{\rm c}$ corresponds to the maximum of $P$. The circle $r=r_{\rm c}$ in the equatorial plane is called the \emph{centre} of the torus.

\subsection{Solution for a polytropic equation of state}

To go further, we assume a polytropic equation of state,
\be 
P = K \epsilon^{1 + 1/n},
\label{eq:polytropic}
\ee
where $K$ and $n$ are two constants, $n$ being the \emph{polytropic index} and $K$ the \emph{polytropic constant}. The total energy density $\epsilon = \rho + \Pi$ is the sum of energy density $\rho$ and internal energy $\Pi$, which in the non-relativistic limit, $\Pi \ll \rho$, reduces to $\epsilon \simeq \rho$. Equation~(\ref{eq:def_enthalpy}) is then readily integrated, yielding
\be 
  H = (n+1) \ln \left( 1 + K \epsilon^{1/n} \right) . 
\label{eq:H_epsilon}
\ee
The surface of the torus is defined by $P=0$. From Eq.~(\ref{eq:polytropic}), this corresponds to $\epsilon = 0$, and from Eq.~(\ref{eq:H_epsilon}) to $H = 0$. Therefore, we may rewrite the first integral of motion (\ref{eq:first_integ}) as
\be 
  H = W - W_{\rm s} , 
\label{eq:first_integ_Ws}
\ee
where the constant $W_{\rm s}$ is the value of the potential $W$ at the torus surface. Denoting by $H_{\rm c}$ and $W_{\rm c}$ the values of $H$ and $W$ at the torus centre, Eq.~(\ref{eq:first_integ_Ws}) implies
\be 
  H_{\rm c} = W_{\rm c} - W_{\rm s}.
\label{eq:H_c}
\ee
Let us introduce the dimensionless potential
\be 
  \omega(r,\theta) \equiv \frac{W(r,\theta) - W_{\rm s}}{W_{\rm c} - W_{\rm s}} . 
\label{eq:omega_r_theta}
\ee
From Eqs.~(\ref{eq:first_integ_Ws}) and (\ref{eq:H_c}), we have
\be
    H = H_{\rm c} \omega . 
\ee
Substituting Eq.~(\ref{eq:H_epsilon}) for $H$, we get
\[
  \ln \left( 1 + K \epsilon^{1/n} \right) = \omega 
  \ln \left( 1 + K \epsilon_{\rm c}^{1/n} \right) , 
\]
where $\epsilon_{\rm c}$ is the energy density at the torus centre. Solving for $\epsilon$, we obtain
\be
  \epsilon = \frac{1}{K^n} \left[(K \epsilon_{\rm c}^{1/n} + 1)^{\omega} - 1 \right]^n.
\label{eq:density}
\ee

At this stage, our torus model is determined by five parameters (in addition to the black hole mass $M$): the Kerr spin parameter $a$, the dimensionless specific angular momentum $\lambda$, the polytropic index $n$, the polytropic constant $K$, and the central density $\epsilon_{\rm c}$. From $\lambda$ and $a$, we evaluate $\ell_0$ via Eq.~(\ref{eq:def_lambda}). The values of $\ell_0$ and $a$ fully determine the potential $W(r,\theta)$ according to formulae (\ref{eq:W})-(\ref{eq:Omega_ell0}). Since we are seeking for a Roche lobe filling torus, the value $W_{\rm s}$ of the potential $W$ at the torus surface must be set to the value at the Roche lobe (cf. Sec.~\ref{s:fluid_torus}): 
\be
  W_{\rm s} = W(r_{\rm in}, \pi/2) . 
\ee 
Given $W_{\rm s}$ and $H_{\rm c}$ (deduced from $\epsilon_{\rm c}$ via (\ref{eq:H_epsilon})), we determine $W_{\rm c}$ by (\ref{eq:H_c}). We know then entirely the dimensionless potential $\omega(r,\theta)$ as given by formula~(\ref{eq:omega_r_theta}). We can thus compute the energy density everywhere in the torus by formula (\ref{eq:density}). 

Note that, by construction [cf. Eq.~(\ref{eq:omega_r_theta})], $\omega$ is zero at the surface of the torus and 1 at the centre. Between $0 \le \omega(r,\theta) \le 1$ the toroidal equipotentials go from the biggest possible, the torus surface, down to a single line, a circle at $r_{\rm c}$. Open equipotential surfaces have $\omega(r,\theta) < 0$. The size of the Roche lobe tori depends on the spin. The higher the spin the narrower is the angular momentum distribution which sets the location of the critical radii, and hence the smaller are the tori. For a fixed value of $a$, the Roche torus is the largest, with an infinite outer radius $r_{\rm out}$, when $\lambda = 1$, i.e., $\ell_0 = \ell_{\rm mb}(a)$.

\section{Adding physics}
\label{sec:physics}

In what follows, we set $n=3/2$, which corresponds to the adiabatic index $\gamma = 1 + 1/n = 5/3$ of a non-relativistic  ideal gas with no radiation pressure. This is consistent for the very optically thin medium.

\subsection{Thermodynamic quantities}

In optically thin gas pressure supported ion tori the radiation pressure can be neglected. The total pressure $P$ is expressed as the sum of the magnetic and gas contributions, $P_{\rm mag}$ and $P_{\rm gas}$:
\be
  P = P_{\rm mag} + P_{\rm gas}
\label{eq: eos}
\ee
The magnetic and gas pressures are assumed to be some fixed parts of the total pressure (an assumption that is often made in analytic models of accretion structures):
\be
  P_{\rm mag} = \frac{B^2}{24\pi} = \beta\,P, ~~~~~ P_{\rm gas} = (1 - \beta)\,P.
\label{eq: p}
\ee
Here, $B$ is the intensity of magnetic field. The gas is assumed to be a two-temperature plasma, with ${\mu_{\rm i}}$, ${\mu_{\rm e}}$, $T_{\rm i}$, $T_{\rm e}$ being the mean molecular weights and temperatures of ions and electrons, respectively. The gas pressure is then expressed as the sum of the ion and electron contributions:
\be
  P_{\rm gas} = P_{\rm i} + P_{\rm e} = \frac{k_{\rm B}}{m_{\rm u}} \epsilon \left( \frac{T_{\rm i}}{\mu_{\rm i}} + \frac{T_{\rm e}}{\mu_{\rm e}} \right),
\label{eq:ion-electron}
\ee
where $k_{\rm B}$ is the Boltzmann constant and $m_{\rm u}$ is the atomic mass unit. Let us write the ion and electron temperatures as follows:
\be
  T_{\rm e} = f(\omega) \mu_{\rm e} \frac{m_{\rm u} P_{\rm gas} }{k_{\rm B} \epsilon}, ~~~~~
  T_{\rm i} = g(\omega) \mu_{\rm i} \frac{m_{\rm u} P_{\rm gas} }{k_{\rm B} \epsilon}.
\ee
where $f$ and $g$ are linear functions of the equipotential function $\omega$.

The condition is that they are equal at the surface: $T_{\rm i} = T_{\rm e}$ when $\omega = 0$, and that they are at some decided ratio $\xi$ at the centre: $\xi\,T_{\rm i} = T_{\rm e}$ when $\omega = 1$. That leads to
\bea
f(0) &=& \frac{\mu_{\rm i}}{\mu_{\rm e} + \mu_{\rm i}} \equiv \cal M, \\
f(1) &=& \frac{\mu_{\rm i} \xi}{\mu_{\rm e} + \mu_{\rm i} \xi} \equiv \cal M_{\xi},
\eea
and since Eq.~(\ref{eq:ion-electron}) is fulfilled when $f(\omega)+g(\omega)=1$, one easily finds $g(0)$ and $g(1)$ which allows to write explicit expressions for the two temperatures,
\bea
  T_{\rm e} &=& \left[(1-\omega) \mathcal{M}  + \omega \mathcal{M}_{\xi}\right] \mu_{\rm e} \frac{(1-\beta) m_{\rm u} P}{k_{\rm B} \epsilon},\\
  T_{\rm i} &=& \left[\frac{\mu_{\rm e}}{\mu_{\rm i}}\mathcal{M} + \omega (\mathcal{M} - \mathcal{M}_{\xi})\right] \mu_{\rm i} \frac{(1-\beta) m_{\rm u} P}{k_{\rm B} \epsilon}.
\label{eq: temperature}
\eea
It is often practical to express equations in terms of the dimensionless temperatures
\be
  \theta_{\rm e} = \frac{k_{\rm B} T_{\rm e}}{m_{\rm e} c^2}, ~~~~ {\rm and} ~~~~
  \theta_{\rm i} =\frac{k_{\rm B} T_{\rm i}}{m_{\rm i} c^2} .
\label{eq: temp}
\ee
Here $m_{\rm e}$ and $m_{\rm i}$ are the electron and ion masses.

\begin{figure}
\includegraphics[width=\linewidth]{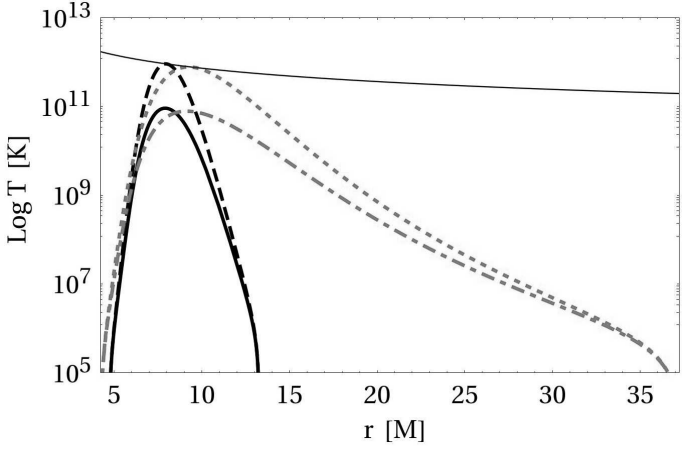}
\caption{Ion (dashed/dotted) and electron (solid/dot-dashed) temperature distribution throughout a $\lambda = 0.3$ torus (black) and a $\lambda = 0.7$ torus (grey) around a Schwarzschild black hole. We assume a central ion to electron temperature ratio of $\xi=0.1$. The thin black line marks $T_{\rm vir}$.}
\label{fig:temp}
\end{figure}
%

Figure~\ref{fig:temp} shows the temperature distribution in tori. It is different from the ADAF self-similar solution \citep[e.g.,][]{ny95} which {\it being self-similar} must be a monotonic function of radius. Here, the non-monotonicity of the temperature curves has its roots in the existence of {\it boundaries}, the outer at the outer torus radius and the inner at the cusp. The ion and electron temperatures have a maximal separation (given by $\xi$) at the torus centre and converge towards either end of the torus. Obviously, the temperature in our model is zero at these boundaries. Note, that in the torus interior our curves and these from \citet{ny95} are comfortably consistent. Using Eq.~(\ref{eq:ion-electron}) one can now derive the relation for the polytropic constant at the torus centre, 
\be
  K = \frac{k T_{\rm e,c}}{(1-\beta) m_{\rm u} \epsilon_{\rm c}^{2/3} \mu_{\rm e}\mathcal{M}_{\xi}}
\ee
where $T_{\rm e,c}$ is the central electron temperature. The total pressure is
\be \label{eq:total_pressure}
  P = \frac{k_{\rm B} T_{\rm e,c}}{(1-\beta)\mu_{\rm e} m_{\rm u} \mathcal{M}_{\xi}} \frac{\epsilon^{5/3}}{\epsilon_{\rm c}^{~2/3}},
\ee
and the magnetic field 
\be
\label{eq:Bfield}
  B = \left[ \frac{24\pi\beta}{(1-\beta)} \frac{k_{\rm B} T_{\rm e,c}}{\mu_{\rm e} m_{\rm u} \mathcal{M}_{\xi}} \frac{\epsilon^{5/3}}{\epsilon_{\rm c}^{~2/3}} \right]^{1/2} .
\ee

The thermodynamic relations between $T$, $P$ and $\epsilon$ require the knowledge of the mean molecular weight $\mu$ linked to the ion and electron molecular weights by: 
\be
  \mu = \left[\frac{1}{\mu_{\rm i}} + \frac{1}{\mu_{\rm e}}\right]^{-1}   \approx \frac{2}{1 + 3X + Y/2}
\ee
with
\bea
  \mu_{\rm i} = \frac{\epsilon}{n_{\rm i} m_{\rm u}} = \frac{4}{4 X + Y}   \quad {\rm and } \quad
  \mu_{\rm e} = \frac{\epsilon}{n_{\rm e} m_{\rm u}} = \frac{2}{1 + Y},
\eea
where $X$ and $Y$ are the hydrogen and helium abundances, assumed to be equal to $X=0.75$ and $Y=0.25$ as in \citet{ny95}. On the one hand this implies the effective molecular weight of ions to be $\mu_{\rm i} = 1.23$ and for electrons $\mu_{\rm e}= 1.14$, and on the other defines the electron and ion number densities, $n_{\rm e}$ and $n_{\rm i}$, with respect to the energy density. In the following, we will give general formulae in which the factor $n_{\rm i}$ is replaced by the sum over ion species,
\be 
  \bar{n}=\sum Z^{2}_j n_j,
\ee
where $Z_j$ and $n_j$ are respectively charge and number density of the $j^{th}$ ion species.

\subsection{Radiative processes}
\label{sec:radproc}

We have to complete the set of equations by specifying the physical processes which will be involved in the radiative cooling. We consider a two-temperature plasma cooled by synchrotron radiation, inverse Compton process and bremsstrahlung emission. A very convenient and general description of such cooling processes has been presented by \citet{ny95}. We closely follow their approximations and procedures, and also use their formulae (correcting one typo). For the reader's convenience and completeness, we give here a summary of these. Electron--positron pair creation and annihilation is neglected but as shown by \citet{bjo+96} and \citet{kus+96} this is justified in most cases of interest.

The following subsections aim at determining the emission and absorption coefficients inside the torus due to bremsstrahlung, synchrotron radiation and Compton processes. The emission coefficient (resp. absorption coefficient), $j_{\nu}$ (resp. $\alpha_{\nu}$), allows to compute the increment (resp. decrement) of specific intensity $I_{\nu}$ when progressing inside the emitting medium by a small distance $\mathrm{d}l$:
\be
\begin{array}{lcl}
  \mathrm{d}I_{\nu} &=& j_{\nu}\,\mathrm{d}l, \\
  \mathrm{d}I_{\nu} &=& -\alpha_{\nu}\,I_{\nu}\,\mathrm{d}l. 
\end{array}
\label{eq:defja}
\ee
The cgs unit of $j_{\nu}$ is $\mathrm{erg}\,\mathrm{cm}^{-3}\,\mathrm{s}^{-1}\,\mathrm{ster}^{-1}\,\mathrm{Hz}^{-1}$ while $\alpha_{\nu}$ is expressed in $\mathrm{cm}^{-1}$.

\subsubsection{Bremsstrahlung}
The rate at which energy is lost due to bremsstrahlung, $f^-_{\rm br} = {\mathrm{d}E_{\mathrm{br}} / \mathrm{d}t\,\mathrm{d}V}$, includes emission from both ion-electron and electron-electron collisions \citep{sve82,ste+83}:
\be 
  f^-_{\rm br} = f^{-}_{\rm ei} +  f^{-}_{\rm ee}. 
\ee 
The ion-electron bremsstrahlung cooling rate is given by 
\be
  f^{-}_{\rm ei}= n_{\rm e} {\bar n} \sigma_{\rm T}\alpha_{\rm f} m_{\rm e} c^3 F_{\rm ei}(\theta_{\rm e})
\label{brei} 
\ee 
where $\alpha_{\rm f}=1/137$ is the fine structure constant and the function $F_{\rm ei}(\theta_{\rm e})$ has the approximate form in units of $[\mathrm{erg}\, \mathrm{s}^{-1} \mathrm{cm}^{-3}]$
\bea
  F_{\rm ei}(\theta_{\rm e}) 
  & = & 4 \left( {2 \,\theta_{\rm e} \over \pi^3} \right)^{1/2} \left( 1+ 1.781\,\theta^{\,1.34}_{\rm e} \right) , 
  ~~~~~~~~~~~~  \theta_{\rm e} < 1 , \nn \\
&&\\
  &=& {9 \,\theta_{\rm e}\over 2 \pi} \left[ \ln (1.123 \,\theta_{\rm e} + 0.48) + 1.5 \right] , 
  ~~~~~~ \theta_{\rm e} > 1 . \nn
\eea
In the original formula quoted by \citet{ste+83} there is a number 0.42 instead of 0.48 \citep[see][]{ny95}.

For the electron--electron bremsstrahlung cooling rate \citet{sve82} gives the following formula 
\be
  f_{\rm ee}^- = n_{\rm e}^2 r_{\rm e}^2 \alpha_{\rm f} m_{\rm e} c^3 F_{\rm ee}(\theta_{\rm e}) ,
\ee
where the function $F_{\rm ee}(\theta_{\rm e})$ is given in units of $[\mathrm{erg}\, \mathrm{s}^{-1} \mathrm{cm}^{-3}]$ and has the approximate form
\bea
  F_{\rm ee}(\theta_{\rm e})
  & = & {20 \over 9 \pi^{1/2}} \left(44 - 3 \pi^2\right) \theta_{\rm e}^{\,3/2} \nn\\
  && \times \left(1 + 1.1\,\theta_{\rm e} + \theta_{\rm e}^2 - 1.25 \,\theta_{\rm e}^{\,5/2}\right) ,
  ~~~~~~~~  \theta_{\rm e} < 1 , \nn \\
&&\\
  & = & 24 \,\theta_{\rm e} \left[ \ln (2\eta\,\theta_{\rm e}) + 1.28 \right] ,
  ~~~~~~~~~~~~~~~~~~~ \theta_{\rm e} > 1 , \nn
\eea
where $r_{\rm e} = e^2/m_{\rm e} c^2$ is the classical radius of electron and the Euler number $\eta = \exp (-\gamma_{\rm E}) = 0.5616$ given by the Euler-Mascheroni constant, $\gamma_{\rm E}$. And again, we replace in the original formula 5/4 with 1.28 \citep[see][]{ny95}. 

With the bremsstrahlung cooling rate given above one can express the bremsstrahlung emission coefficient as 
\be
j^{\rm \,br}_{\nu}  = f^-_{\rm br} \,\frac{1}{4 \pi} \frac{h}{k_{\rm B} T_{\rm e}} \exp \left( -\frac{h \nu}{k_{\rm B} T_{\rm e}} \right) \bar{G},
\ee
where $h$ is the Planck constant, the $1/4\pi$ factor assumes isotropic emission in the emitter's frame and $\bar{G}$ is the velocity averaged Gaunt factor given by \citet{ryb+86}:
\bea
   \bar{G} & = & \left( \frac{3}{\pi} \frac{k_{\rm B} T_{\rm e}}{h \nu} \right)^{1/2} , ~~~~~~~~~~~~~~~~~~~~~~~~~~~~~~~~~~~~~~~ \frac{k_{\rm B} T_{\rm e}}{h \nu} < 1 , \nn\\
  && \\
  & = & \frac{\sqrt{3}}{\pi} \ln \left( \frac{4}{\gamma_{\rm E}} \frac{k_{\rm B} T_{\rm e}}{h \nu} \right) ,  ~~~~~~~~~~~~~~~~~~~~~~~~~~~~~~ \frac{k_{\rm B} T_{\rm e}}{h \nu} > 1 . \nn
\eea

\subsubsection{Synchrotron cooling}
The emission coefficient for synchrotron emission by a relativistic Maxwellian distribution of electrons writes \citep[see][]{pac70}
\be
  j_{\nu}^{\rm \,sy} =  \frac{1}{4\pi} \frac{e^{2}}{c\sqrt{3}} \frac{4\pi n_{\rm e}\nu}{K_2(1/\theta_{\rm e})} \, M(x_{\rm M}) ,
\label{eq:synch}
\ee
with a factor $1/4\pi$ again for isotropic emission in the emitter's frame and the fitting function,  
\bea
  M(x_{\rm M}) &= &\frac{4.0505 \alpha}{x_{\rm M}^{1/6}} \left( 1 + {0.40 \beta \over x_{\rm M}^{1/4}} + {0.5316 \gamma \over  x_{\rm M}^{1/2}} \right) \, \exp\left(-1.8899x_{\rm M}^{1/3}\right) , \nn \\
  & & 
\eea
where
\be
  x_{\rm M} = \frac{2\nu}{3\nu_0 \,\theta_{\rm e}^2}, ~~~ \nu_0 = \frac{eB}{2\pi m_{\rm e} c}
\label{eq:xM}
\ee
and the parameters $\alpha$, $\beta$ and $\gamma$ are tabulated for a range of temperatures in
\citet{mah+96}.
The fitting formula is valid only for $\theta_{\rm e} \gtrsim 1$, i.e. $T_{\rm e} \gtrsim 10^{8}$ K, which is satisfied for applications to ion tori because the synchrotron emission that dominates the central regions of a torus comes, like in other advective flows, from relativistic electrons in the tail of the Maxwellian distribution. At lower temperatures or in the outer torus regions emission is dominated by bremsstrahlung. Below a critical frequency, $\nu_{\rm c}$, the synchrotron spectrum becomes self-absorbed. As the flow in ion tori is very much akin to a spherical flow, this frequency can be obtained by equating the synchrotron emission in a sphere of some radius, $R$ [cm], to the Rayleigh-Jeans blackbody emission from the surface of that sphere and solving for $x_{\rm M}$ which is then substituted into 
\be
\nu_{\rm c} = \frac{3}{2} \nu_0 \,\theta_e^2 x_{\rm M} .
\ee
We assume that at low frequencies, $\nu < \nu_{\rm c}$, the absorption is locally given by Kirchhoff's law, i.e., the low-frequency part of the synchrotron emission behaves like a blackbody.

\subsubsection{Compton cooling}
The soft bremsstrahlung and synchrotron photons in an ion torus filled with a thermal distribution of electrons are (inverse) Compton scattered to higher energies. Especially in the central regions of the flow this can be an important cooling mechanism. There is a probability $\mathcal{P}$ that a seed photon of some initial energy, $E_{\rm in} = h \nu$, is in optically thin material scattered to an amplified energy $E_{\rm out} = \mathcal{A} E_{\rm in}$, where
\be
  \mathcal{P}  =  1 - \exp{(-\tau_{\rm es})} , 
\qquad \mathrm{and} \qquad 
  \mathcal{A}  =  1 + 4\theta_{\rm e} + 16 \theta^2_{\rm e} .
\ee
Thus, the energy exchange between electrons and photons depends only on the electron temperature $\theta_{\rm e}$ and the probability that a photon will interact with an electron, which is given by the electron scattering optical depth $\tau_{\rm es} = \int{n_{\rm e} \, \sigma_{\rm T} \, \mathrm{d}l }$. \citet{der+91} and \citet{esi+96} derived an approximate prescription for the energy enhancement factor due to Compton scattering, which is defined as the average energy change of a seed photon:
\be
  \eta  =  1 + \eta_1 + \eta_2\left({x \over \theta_{\rm e}} \right)^{\eta_3},
\ee
where
\bea
  \eta_1  &=& {\mathcal{P}(\mathcal{A}-1) \over 1 - \mathcal{PA}}, \nn \\ 
  \eta_2  &=&  3^{- \eta_3} \eta_1 , \\
  \eta_3  &=&  -1 - \ln \mathcal{P}/\ln \mathcal{A} .\nn
\eea
The dimensionless energy is given by $x  =  {h \nu \over m_{\rm e} c^2} $. As the emerging photons cannot gain more energy than the electrons they collide with have, there is an upper limit on the energy, $x \lesssim 3 \, \theta_{\rm e}$. Note that in our simple ion torus scenario we apply, without loss of generality, the Thomson cross section rather than the  Klein-Nishina cross section.

Comptonised emission is $\eta - 1$ times the seed photon distribution. The part of the spectrum which can be Comptonised lies between the critical synchrotron self-absorption edge, $x = x_{\rm c} = h \nu_{\rm c} / m_{\rm e} c^2$, and $x = 3 \, \theta_{\rm e}$. 

{\it Comptonisation of bremsstrahlung emission} is then given by
\bea
  j^{\rm \, br, C}_{\nu} = j^{\rm \,br}_{\nu} \, 3 \eta_1 \,\theta_{\rm e}\left\{\left({1 \over 3} -{x_{\rm c} \over 3\theta_{\rm e}}\right)
        - {1 \over \eta_3 +1} \left[\left({1 \over 3}\right)^{\eta_3 + 1}
        - \left({x_{\rm c} \over 3 \theta_{\rm e}}\right)^{\eta_3  + 1}
        \right] \right\}. \nn\\
\eea
Note, that the corresponding expression in \citet{ny95} has a typo which we corrected here.

{\it Comptonisation of synchrotron radiation} which is emitted mostly at the self-absorption frequency, $\nu_{\rm c}$, is given by
\be
 j^{\rm \, sy, C}_{\nu} = j^{\rm \, sy}_{\nu} \left[\eta_1 - \eta_2
                           \left({x_{\rm c} / \theta_{\rm e}}\right)^{\eta_3}
                           \right] .
\ee
We calculate Comptonisation to second order and assume that photons which are up-scattered to $\theta_{\rm e}$ form a Wien tail.

\subsubsection{Total cooling}

The total emission coefficient is the sum of all radiative contributions:
\be
  j_{\nu} = j^{\rm br}_{\nu} + j^{\rm br, C}_{\nu} + j^{\rm sy}_{\nu} + j^{\rm sy, C}_{\nu} .
\label{eq:emissivity}
\ee
For a medium in local thermodynamic equilibrium (LTE) at temperature $T$, the emission coefficient (\ref{eq:emissivity}) and the absorption coefficient are related by means of Kirchhoff's law: 
\be
\alpha_{\nu} = \frac{j_{\nu}}{B_{\nu}(T)}
\ee
where $B_{\nu} $ is Planck's law of blackbody radiation. For the typical temperatures of the ion tori considered in this paper,  $\alpha_{\nu}$ is negligible, thus the absorption can be safely ignored. 
 
\begin{figure}[h!]
\includegraphics[width=\linewidth]{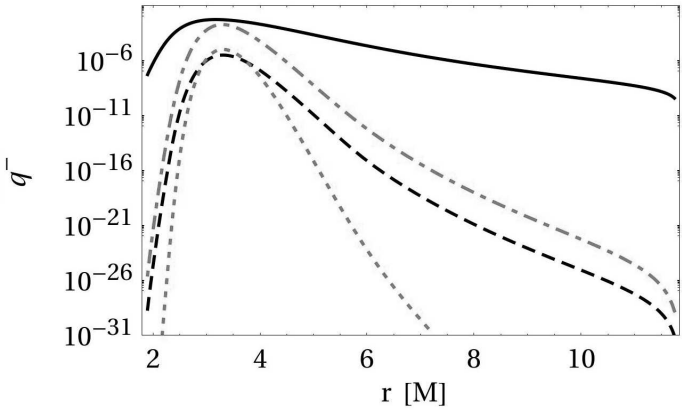}
\caption{Radiative cooling rates of a $\lambda$ = 0.7 torus around a spin $a_*$ = 0.9 black hole. The contributions from bremsstrahlung (black, dashed), synchrotron (black, solid), Comptonised bremsstrahlung (grey, dot-dashed) and Comptonised synchrotron emission (grey, dotted) are shown as a function of radius.}
\label{fig:cooling}
\end{figure}

Integration of $j_{\nu}$ over the whole frequency range gives the total cooling rate $q^-$ at each point $(r,\theta)$ in the torus. Fig.~\ref{fig:cooling} shows the different contributions to the total cooling rate from the equatorial plane, $\theta = \pi/2$, of a $\lambda$ = 0.3 torus.

%

Radiation that originates close to a black hole is influenced by various relativistic effects, such as the gravitational bending of light rays, gravitational redshift and Doppler beaming. We ray-trace the emission derived in Eq.~(\ref{eq:emissivity}) from each point (r, $\theta$) inside the Roche lobe equipotential back to the observer explicitly including all relativistic effects in light propagation. This is described in the next two sections.

\section{Ray-tracing}
\label{sec:ray-tracing}
We use the General relativitY Orbit Tracer of the Observatory of Paris\footnote{Freely available at the  URL \texttt{http://gyoto.obspm.fr}} (GYOTO), presented in \citet{vin+11}. GYOTO allows integrating null and timelike geodesics in any analytical or numerical metric. It is optimised to compute geodesics in Kerr spacetime efficiently, and is able to integrate  the radiative transfer equation in optically thin media.

The required inputs for GYOTO are the values of the emission coefficient, $j_{\nu}$, and absorption coefficient, $\alpha_{\nu}$, at any point inside the torus. These quantities, defined in section~\ref{sec:radproc}, are known analytically for the ion torus, so the integration is straightforward. In particular, the absorption coefficient is $0$ everywhere.

Photons are launched from the observer's screen, supposed to be spatially at rest at $r = 100 M$, in a given specified solid angle. The null geodesics are integrated until the torus is reached (or until the photon escapes too far from the torus) and the specific intensity $I_{\nu}$ is updated at each step inside the torus by using Eq.~(\ref{eq:defja}). A map of specific intensity is thus obtained for a given set of directions on sky. It is then straightforward to compute the observed specific flux $F_{\nu}$ by summing these values of specific intensities over all directions of incidence.

Two kinds of GYOTO-computed quantities are presented below. The spectra (Figs.~\ref{fig:spectra_spin} -~\ref{fig:spectra_inc}) show the quantity $\nu\,F_{\nu}$ for different values of frequencies. The images (Fig.~\ref{fig:images}) are maps of specific intensity $I_{\nu}$.

Let us stress that the whole C++ code for the ion torus is now included in the open source GYOTO code, available at the above mentioned URL.

\section{The spectra and images of ion tori}
\label{sec:results}
This section presents the spectra and images of an ion torus surrounding a Kerr black hole. Table~\ref{tab:setup} gives the reference values of the parameters used for the computations. The value chosen for the central density $\epsilon_{\rm c}$ is standard for Sgr~A*~\citep[see e.g. the values in][]{yua+03, liu+04, mos+09, dod+10}. Various parameters are varied in the computations: the spin, $a$, the dimensionless angular momentum, $\lambda$, the magnetic to total pressure ratio, $\beta$, the electron to ion temperature ratio, $\xi$, the ratio between central electron temperature and virial temperature, and the inclination, $i$.

The following sections~\ref{sec:spectra} and~\ref{sec:image} show the impact of these parameters on the spectrum and image of the ion torus.

\begin{table}[ht]
\center
\begin{tabular}{l l}
\hline 
\hline
&\\
parameter & value\\
&\\
\hline 
\hline
&\\
$a$                                   & $0.5 M$\\
$\lambda$                             & 0.3 \\
$\beta$                               & 0.1 \\
$n$                                   & 3/2 \\
$\epsilon_{\rm c}$ [$\mathrm{g} \, \mathrm{cm}^{-3}$]   & $10^{-17}$ \\
$\xi$                                 & 0.1 \\
$T_{\rm e,c}$                         & 0.02 $T_{\rm vir}$ \\
$i$ [ $^{\circ}$]                     & 80\\
&\\
\hline
\hline 
\end{tabular}
\caption[Ion torus set-up]{Parameters for our reference ion torus. Parameter $i$ is the inclination angle of the black hole rotational axis towards the line of sight. $T_{\rm vir}$ is the virial temperature.}
\label{tab:setup}
\end{table}

\subsection{Torus spectra}
\label{sec:spectra}

\begin{figure}[h!]
\center
 \includegraphics[width=\linewidth]{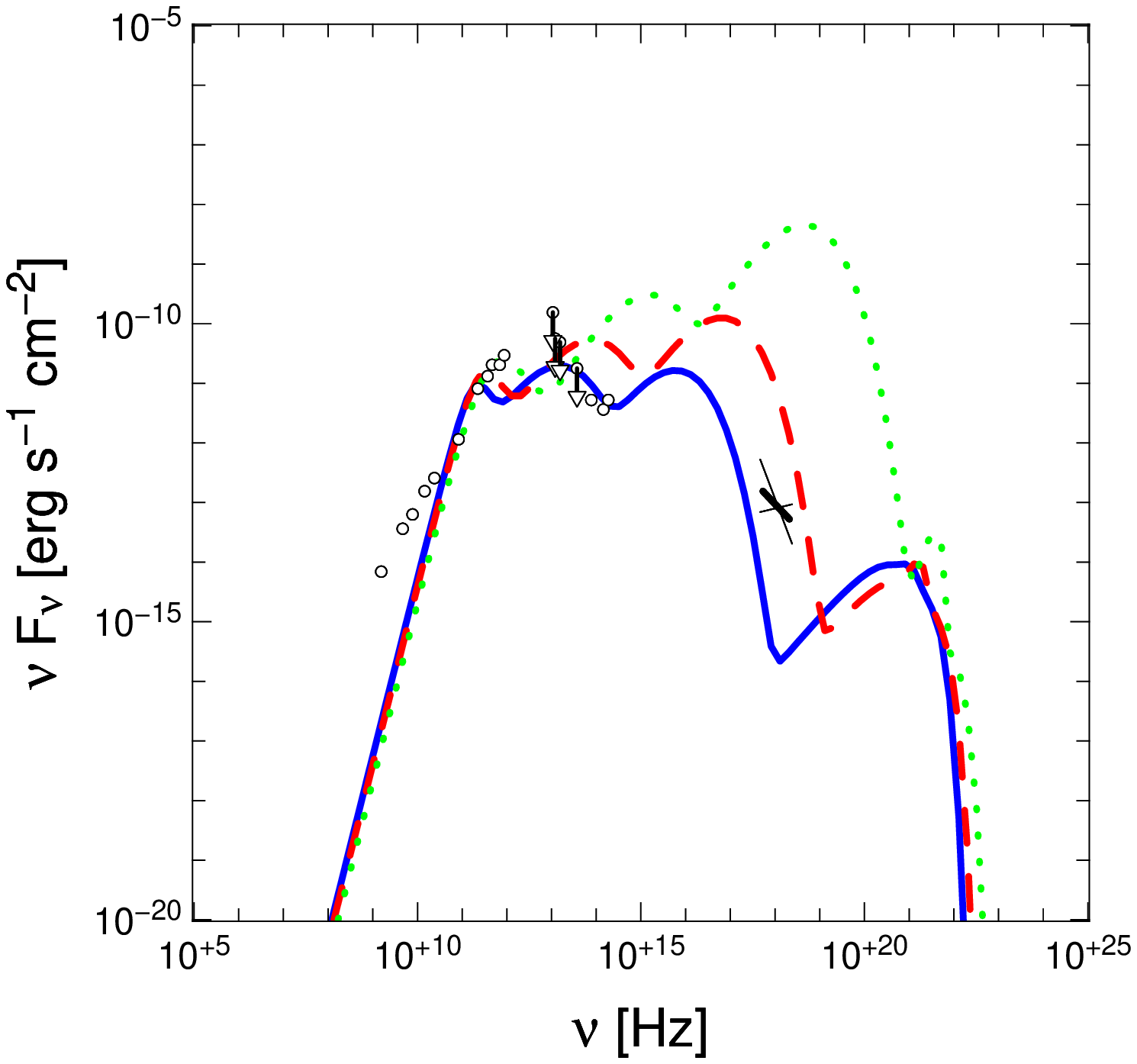}
 \caption{\textbf{Impact of spin} on an ion torus spectrum: $a = 0$ (solid blue), $0.5M$ (dashed red) or $0.9M$ (dotted green). All other parameters are set to their reference values listed in Tab.~\ref{tab:setup}. The black observed data are taken from \citet{zha+01}, \citet{zyl+95}, \citet{mar+08} for radio and sub-mm data, \citet{tel+96}, \citet{cot+99}, \citet{eck+06}, \citet{sch+07} for far- and mid-infrared data, \citet{gen+03} for near-infrared data and \citet{bag+03} for the X-ray bow tie. The down pointing arrows refer to upper values. Note that we do not present fitted spectra here.}
 \label{fig:spectra_spin}
\end{figure}

\begin{figure}[h!]
\center
 \includegraphics[width=\linewidth]{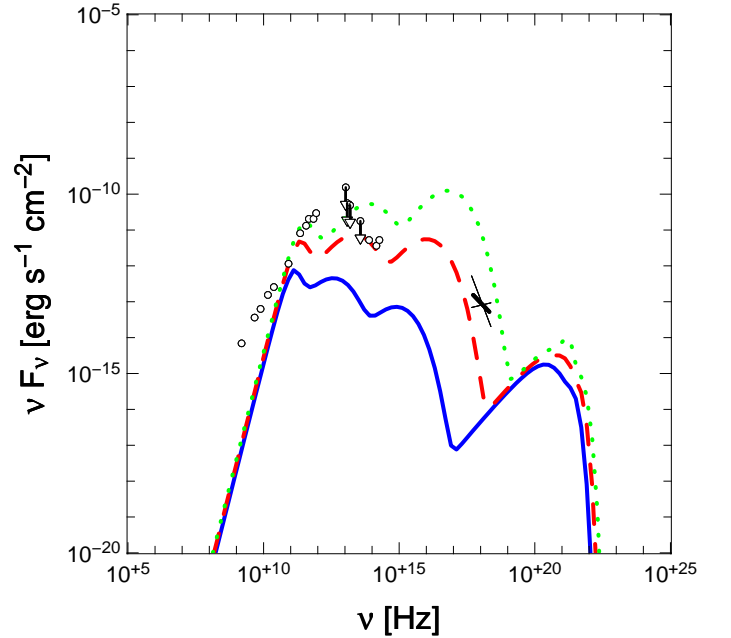}
 \caption{\textbf{Impact of central temperature}: $T_{0}/T_{\mathrm{vir}} = 0.01 $ (solid blue), $0.015$ (dashed red) or $0.02$ (dotted green).}
 \label{fig:spectra_t0}
\end{figure}

\begin{figure}[h!]
\center
 \includegraphics[width=\linewidth]{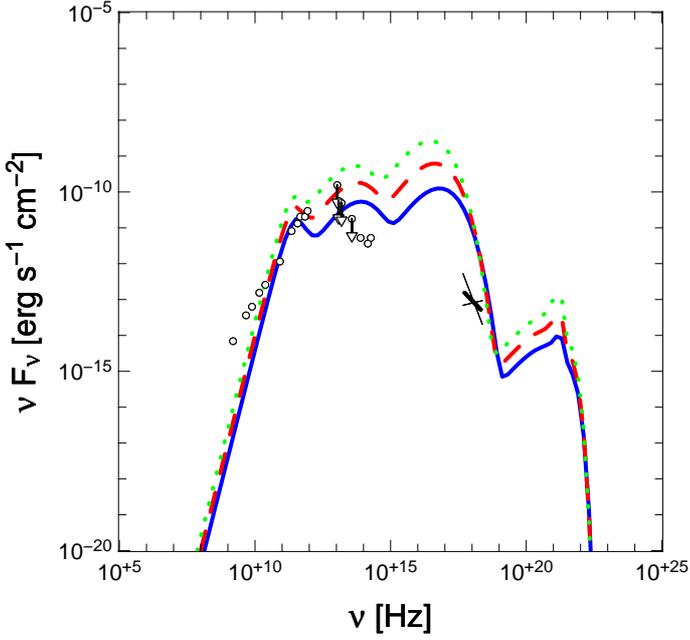}
 \caption{\textbf{Impact of angular momentum}: $\lambda = 0.3 $ (solid blue), $0.45$ (dashed red) or $0.6$ (dotted green).}
 \label{fig:spectra_lambda}
\end{figure}

\begin{figure}[h!]
\center
 \includegraphics[width=\linewidth]{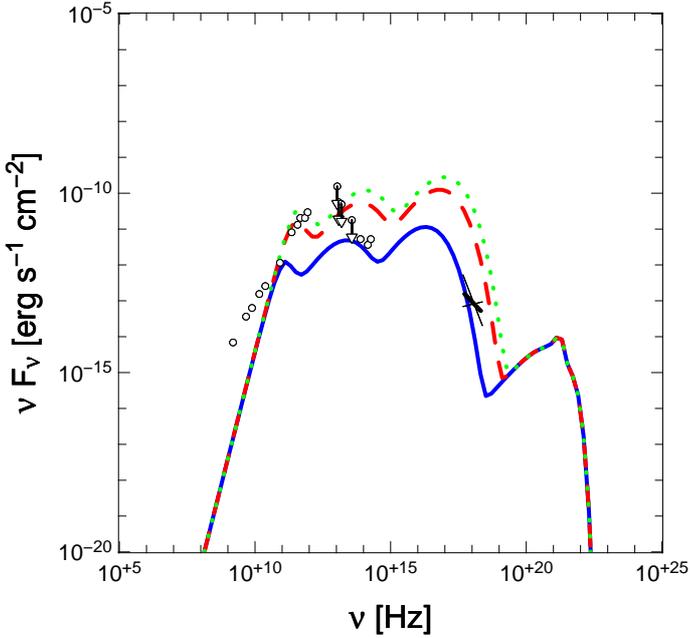}
 \caption{\textbf{Impact of pressure ratio}: $\beta = 0.01 $ (solid blue), $0.1$ (dashed red) or $0.2$ (dotted green).}
 \label{fig:spectra_beta}
\end{figure}

\begin{figure}[h!]
\center
 \includegraphics[width=\linewidth]{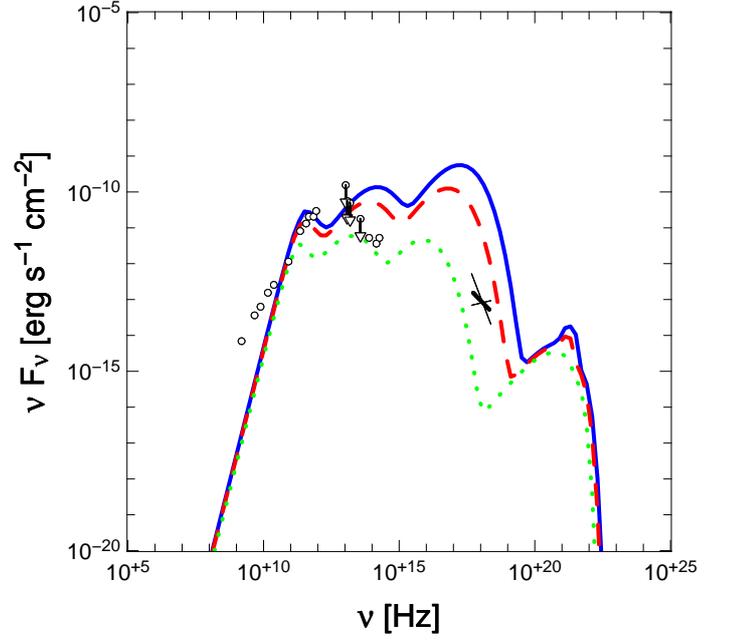}
 \caption{\textbf{Impact of temperature ratio}: $\xi = 0.08 $ (solid blue), $0.1$ (dashed red) or $0.2$ (dotted green).}
 \label{fig:spectra_xi}
\end{figure}

\begin{figure}[h!]
\center
 \includegraphics[width=\linewidth]{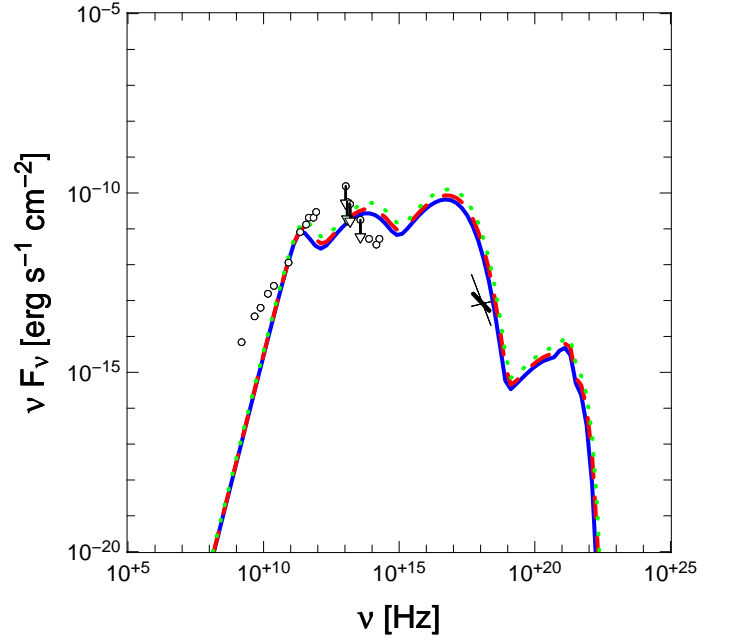}
 \caption{\textbf{Impact of inclination}: $i=40^{\circ}$ (solid blue), $60^{\circ}$ (dashed red) or $80^{\circ}$ (dotted green).}
 \label{fig:spectra_inc}
\end{figure}

Figures~\ref{fig:spectra_spin} to~\ref{fig:spectra_inc} show the impact of various parameters on the observed spectrum of an ion torus. All parameters that have not explicitly specified values are set according to Tab.~\ref{tab:setup}. These figures show that the ion torus model is able to account for the general features of the observed data, by appropriately tuning the various parameters. This includes the X-ray flare ``bow tie'' which in the ion torus model may originate from soft photons that are inverse Compton scattered by the same population of hot electrons that is responsible for the synchrotron emission. Only the flattening of the spectrum at low frequencies is never matched, which is due to the absence of a non-thermal electron distribution in our model, as demonstrated by~\citet{yua+03}. A more evolved model taking into account this effect will thus be developed in future work. 

Figure~\ref{fig:spectra_spin} shows that different spins significantly alter the emission. This is due to the fact, that at high black hole spins the accretion torus shrinks and moves very close to the black hole. Therefore, the flux is being shifted to slightly higher energies. Increasing spin enhances synchrotron flux, while bremsstrahlung is being softened.

Figure~\ref{fig:spectra_t0} shows that increasing the central temperature implies a displacement of the whole spectrum towards higher fluxes. 
The same effect is obtained by increasing the dimensionless angular momentum $\lambda$
(Fig.~\ref{fig:spectra_lambda}). This is due to the fact that the torus puffs up like a balloon when $\lambda$ increases (see Fig.~\ref{fig:equipotential}), thus providing more flux.

Figures~\ref{fig:spectra_beta} and~\ref{fig:spectra_xi} show that increasing the pressure ratio $\beta$ or temperature ratio $\xi$ has an opposite impact: a higher $\beta$ gives a higher synchrotron flux while a higher $\xi$ gives a smaller flux. This is due to the fact that increasing $\beta$ increases the magnetic field strength, while increasing $\xi$ decreases $\mathbf{B}$ (see Eq.~\ref{eq:Bfield}). 

Figure~\ref{fig:spectra_inc} shows that the inclination has little effect on the spectrum. Although the number of illuminated pixels is smaller at higher inclination, the beaming effect is more important. These two phenomena have opposite effect, and the resulting spectrum is not changed much.

Future work will be devoted to studying the possibility of constraining the various parameters, in particular Sgr~A*'s spin, by fitting such spectra to observed data.

%

%

%

%

%

%

\subsection{Torus images}
\label{sec:image}

Each image is a superimposition of a first-order image (the thick distorted ring) and higher order images (the circles centred around the black hole) which result from photons that swirl around the black hole before reaching the observer. There is also a very fine circle of light in each image. It consists of photons that come from a location just outside the photon orbit~\citep[see e.g.][for a definition]{bar+72}. Photons which escape from the region inside the photon orbit are so severely red-shifted that they create an area of reduced intensity on the observer's screen: this is the so-called ``black hole silhouette'' \citep[also known as black hole shadow][]{fal+00}.

\begin{figure*}[h!]
\center
\includegraphics[width=0.48\linewidth]{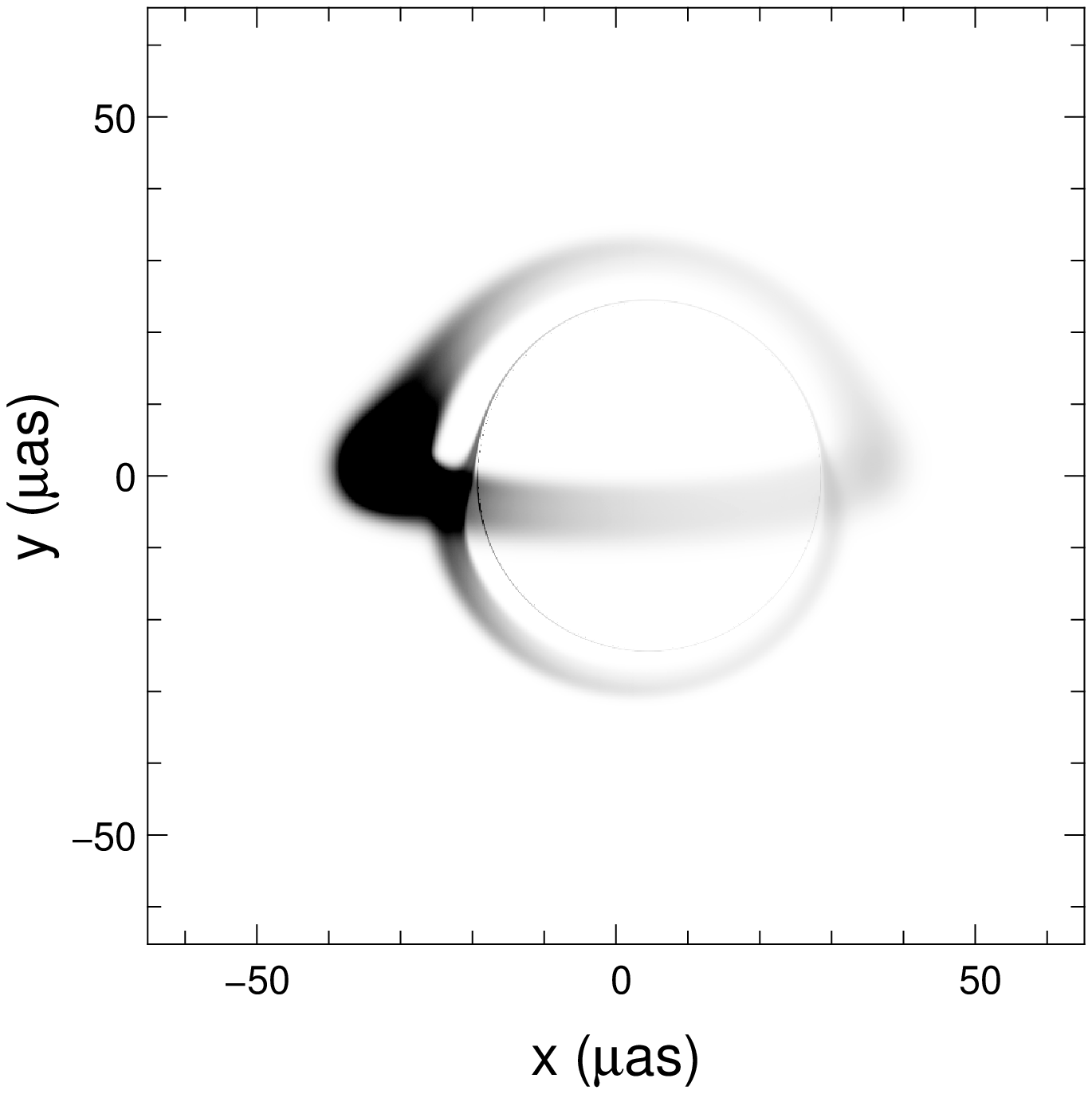}
\hfill
\includegraphics[width=0.48\linewidth]{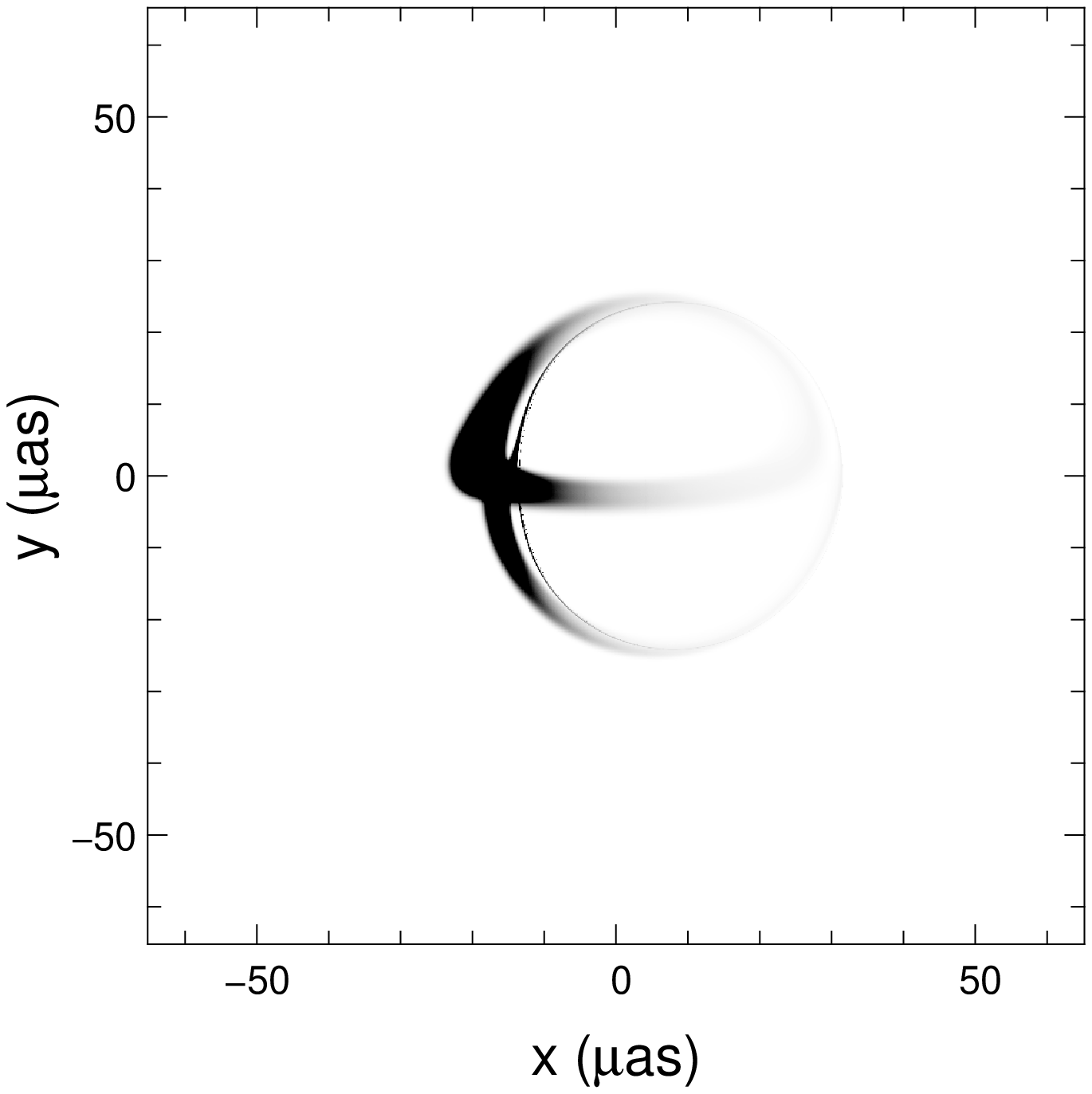}
\includegraphics[width=0.48\linewidth]{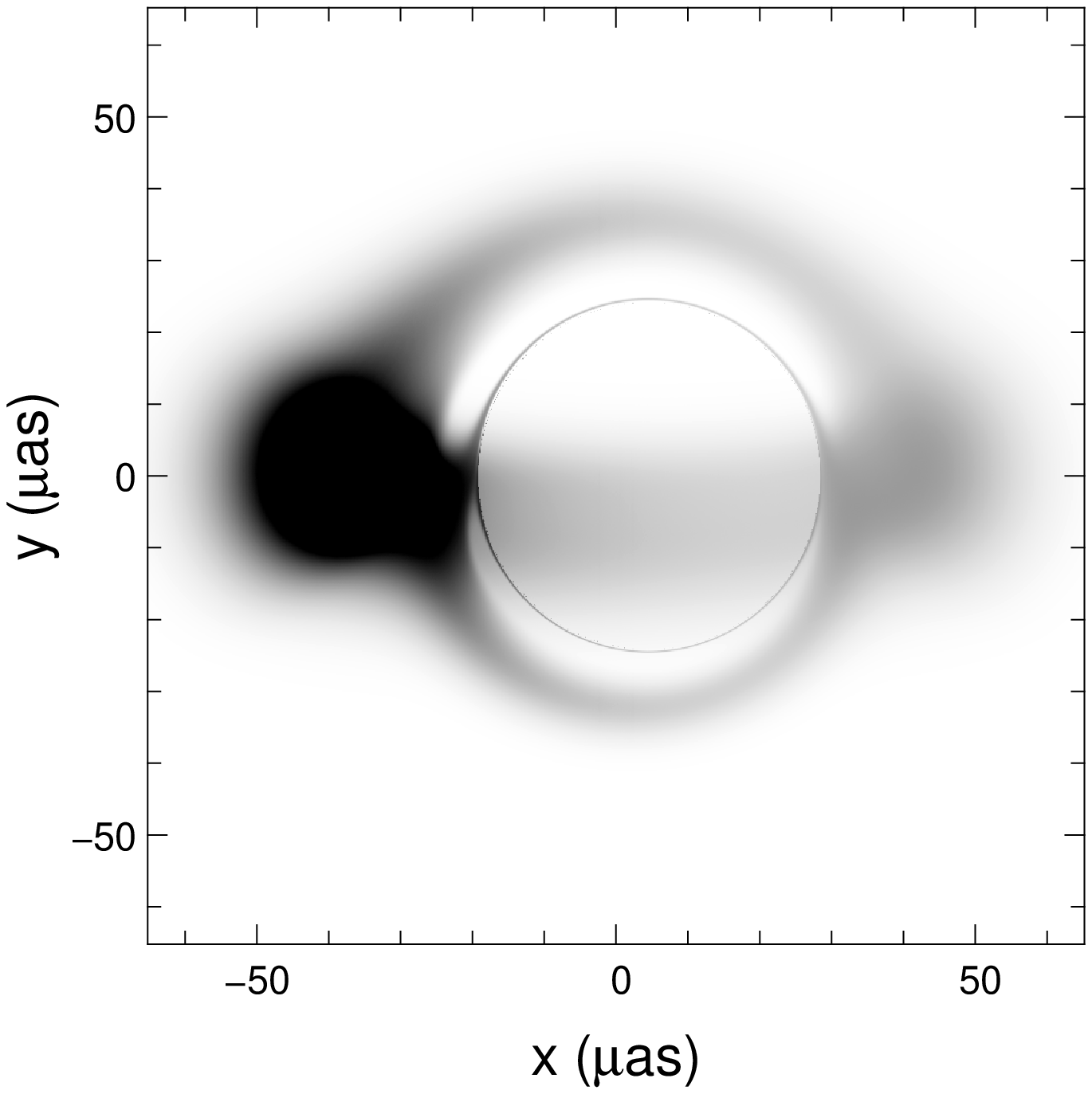}
\hfill
\includegraphics[width=0.48\linewidth]{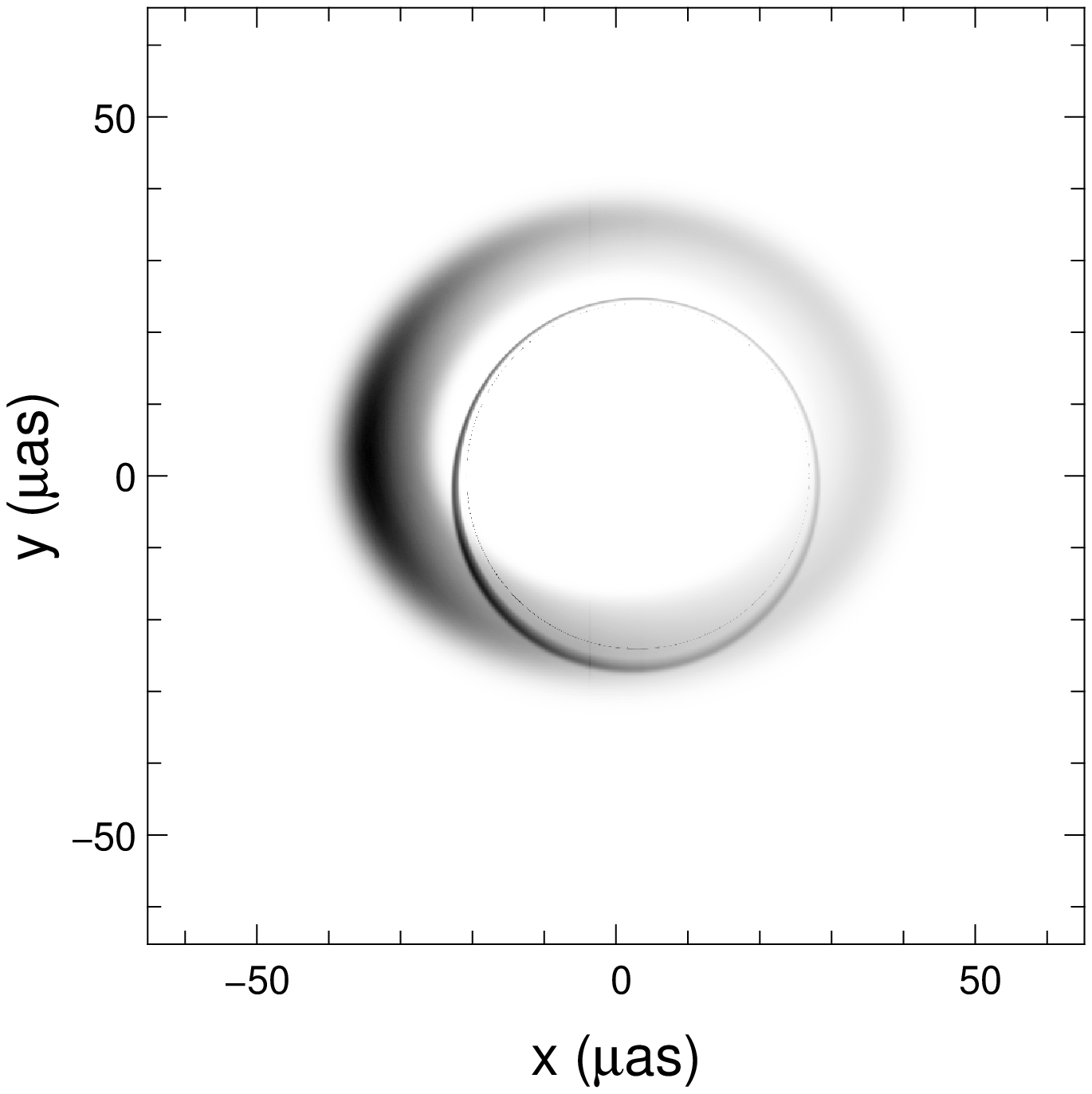}
 \caption{\emph{Upper left:} Image of the reference ion torus corresponding to the parameters listed in Tab.~\ref{tab:setup}, as observed by an observer on Earth, in inverted colours (the darker, the more luminous). 
 The display is intentionally saturated for a better rendering of the less luminous parts of the image.
 \emph{Upper right:} same as in the upper left but with the spin parameter increased to 
 $a=0.9M$. \emph{Lower left:} same as in the upper left but with dimensionless angular momentum 
 increased to $\lambda=0.6$. \emph{Lower right:} same as in the upper left but with the inclination angle decreased to $i=40^{\circ}$.} 
 \label{fig:images}
\end{figure*}
%

The interest of such images is the following. Knowing the particular extent and shape of the accreting region from observations allows to constrain various parameters of the flow. In particular the location and shape of the photon orbit contains information on the geometry of spacetime. Increasing spin displaces the thin photon ring in the image plane off centre and $a_* \gtrsim 0.9$ introduces in addition distortion. Black hole spin, torus dimensionless angular momentum and inclination have each a huge and characteristic impact on the observed image of the torus as depicted in Fig.~\ref{fig:images}. This is also revealed in the spectra Fig.~\ref{fig:spectra_spin} and \ref{fig:spectra_lambda}. The inclination angle, however, does not influence the spectra in a commensurate way (Fig.~\ref{fig:spectra_inc}).

The required technique of measurement of the black hole silhouette is already available thanks to recent progress in millimetre Very Long Baseline Interferometry (mm-VLBI) \citep[see][]{doe+08}. With the near-future sub-mm VLBI instruments direct observation of processes in the range between 5-40 Schwarzschild radii will become feasible and with it also the potential to image radiation from an accretion structure and, possibly, to deduce the existence of an event horizon.

\section{Conclusion and perspectives}
\label{sec:discussion}
We have calculated electromagnetic spectra and images of accretion structures around the central black hole in Sgr~A* using a simple analytic model of ``ion tori'' very similar to these described in the well-known paper by \citet{ree+82}. Results depend on observationally unknown tunable parameters of the model, in particular the black hole spin. The hope is that fitting these analytic models to observations could {\it practically} restrict the allowed parameter range, so that the future sophisticated MHD simulations of Sgr~A* \citep[similar to recent and most advanced ones by][]{dib+12} will be more targeted.

One interesting question here is how accurate are the simple analytic models in comparison with advanced MHD simulations. We are working on a comparison of our spectra and images with these calculated by \citet{dib+12}. If the preliminary impression that these images and spectra are indeed {\it very} similar would survive the test, then it would be much easier to fit theory to observations, and in particular to eventually {\it measure} the black hole spin in Sgr~A*.

We plan to expand our models of ion tori to include a more complete coverage of the parameter space in connection with a ``general'' angular momentum distribution \citep[as in][]{qia+09} and to include, directly and analytically, large scale magnetic fields \citep[as in][]{kom06}. We also plan to calculate models of ion tori
in a non-Kerr background \citep[as advocated by Psaltis and collaborators, e.g.,][]{psa+08,joh+10} to possibly constrain gravity in the strong-field regime of gravity theories alternative to Einstein's general relativity. In the context of flares it may also be interesting to examine non-stationary ion tori \citep[as in][]{abr+83}.

\begin{acknowledgements}
We gratefully acknowledge support from the Polish grants NN203 383136 (OS) and NCN 2011/01/B/ST9/05439 (MAA), the stipends from the Swiss National Science Foundation (OS) and the R\'egion Ile-de-France (FHV) as well as the Swedish VR and the Czech CZ.1.07/2.3.00/20.0071 ``Synergy'' grants (MAA) and support from the French-Polish LEA Astro-PF programme. 
\end{acknowledgements}

\bibliographystyle{aa}
\bibliography{ion_torus}

\end{document}